\documentclass[prd,aps,preprint,showkeys,lengthcheck,twocolumn,nofootinbib,notitlepage,floatfix,superscriptaddress]{revtex4-2}
\usepackage{graphics,graphicx}
\usepackage{amsmath, amssymb}
\usepackage{multirow}
\usepackage{color}
\usepackage{verbatim}
\usepackage{hyperref}
\usepackage[normalem]{ulem}  
\usepackage{ulem}
\usepackage{color}
\usepackage{cancel}
\usepackage{mathtools}
\usepackage{epstopdf}
\usepackage[utf8]{inputenc}
\usepackage[dvipsnames]{xcolor}
\usepackage[normalem]{ulem}

\renewcommand\sout{\bgroup\color{blue} \ULdepth=-.5ex \ULset}

\newcommand{\der}{\text{d}}
\newcommand{\ud}{\mathrm{d}}

\newcommand{\ccBar}{\ensuremath{\text{c}\overline{\text{c}}}}

\def\slashchar#1{\setbox0=\hbox{$#1$}  
\dimen0=\wd0     
\setbox1=\hbox{/} \dimen1=\wd1  
\ifdim\dimen0>\dimen1   
\rlap{\hbox to \dimen0{\hfil/\hfil}} 
#1     
\else     
\rlap{\hbox to \dimen1{\hfil$#1$\hfil}} 
/      
\fi}

\usepackage{lineno}

\begin{document}

\title{Emergence of New Systematics for Open  Charm Production in High Energy Collisions}

\date{\today}
\author{Peter Braun-Munzinger}
    \affiliation{Research Division and ExtreMe Matter Institute EMMI, GSI Helmholtzzentrum f\"{u}r Schwerionenforschung GmbH, Darmstadt, Germany}
  \affiliation{Physikalisches Institut, Ruprecht-Karls-Universit\"{a}t Heidelberg, Heidelberg, Germany}
  \affiliation{Institute of Particle Physics and Key Laboratory of Quark and Lepton Physics (MOE), Central China Normal University, Wuhan 430079, China}
\author{Krzysztof Redlich}
\affiliation{Institute of Theoretical Physics, University of Wroc\l aw, plac Maksa Borna 9, PL-50204 Wroc\l{}aw, Poland}
\address{Polish Academy of Sciences PAN, Podwale 75, 
PL-50449 Wroc\l{}aw, Poland}
\author{Natasha Sharma}
\email{natasha.sharma@cern.ch}
\affiliation{Indian Institute of Science Education and Research (IISER) Berhampur, Ganjam, Odisha, India-760003}
\author{Johanna Stachel}
 \affiliation{Physikalisches Institut, Ruprecht-Karls-Universit\"{a}t Heidelberg, Heidelberg, Germany}

\begin{abstract}
\noindent
 We present the production systematics of open charm hadron yields in high-energy collisions and their description based on the Statistical Hadronization Model of charm  (SHMc). 
 The rapidity density of $D^0, D^+, D^{*+}, D_s^+$ mesons and $\Lambda_c^+$ baryons in heavy ion and proton-proton collisions is analyzed for different collision energies and centralities.  
 The SHMc  is extended to open charm production in 
 minimum-bias and high-multiplicity pp collisions. 
{In this context, we use the link established in  [1,2],  between the rapidity density of open charm hadron yields, $dN_i/dy$,  and the rapidity density of charm-anticharm quark pairs, $dN_{c\bar c}/dy$. 
We demonstrate 
that, in pp, pA and AA collisions,  $dN_i/dy$ scales in leading order with   $dN_{c\bar c}/d\eta$  and for open charm mesons, $D^0,D^+$ and $D^{*+}$ the slope coefficient is quantified by the appropriate thermal density ratio calculated in the SHMc at the chiral crossover temperature, $T_c=156.5$ MeV. The slope coefficient for $dN_{\Lambda_c^+}/dy$ differs at $T_c$ by a factor of $1.97\pm 0.14$ which is attributed to missing charmed-baryon resonances in the PDG.   }
It is also shown that $dN_i/dy$ exhibits power-law scaling with the charged-particle pseudo-rapidity density in high energy collisions and within uncertainties. Furthermore, presently available data on different ratios of open charm
 rapidity densities in high-energy collisions are independent of collision energy and system size, as expected in the SHMc.
\end{abstract}
 \maketitle

\section{Introduction}

The production in relativistic nuclear collisions 
of hadrons with charm and beauty quantum numbers is a key area of focus in the effort to characterize the quark-gluon plasma (QGP). For recent reviews see
{~\cite{Rapp:2008tf,Braun-Munzinger:2015hba,Apolinario:2022vzg,Gross:2022hyw,ALICE:2022wpn,Dong:2019byy,Dong:2019unq,Dong:2019byy,Dong:2019unq}. }
We consider the Statistical Hadronization Model for charm (SHMc) 
\cite{BraunMunzinger:2000px,Andronic:2003zv,Andronic:2006ky,Andronic:2017pug} with special emphasis on its application to open charm hadron production in high-energy proton-proton (pp), proton-nucleus (pA), and nucleus-nucleus (AA)  collisions. 
This model successfully quantifies data on heavy flavor production in heavy ion collisions. In particular, the SHMc has been used to predict the energy dependence of charmonium suppression and enhancement through formation at the QCD phase boundary in the hadronization of the QGP \cite{Andronic:2017pug}. We note here that this approach is essentially parameter-free as it is based on the total open charm cross-section and on knowledge of the mass spectrum of hadronic states with charm and beauty, see ~\cite{Andronic:2017pug}.
The production of charmonia or charmonium-like states and open charm states for different colliding nuclei and centralities has recently been investigated within SHMc \cite{Andronic:2021erx}. By linking the SHMc with the 
(3+1)-dimensional relativistic hydrodynamical expansion of hot and dense matter in ultra-relativistic nuclear collisions, a good description of the transverse momentum distributions of open and hidden charm hadrons 
has been achieved \cite{Andronic:2019wva,Andronic:2021erx,Andronic:2023tui}. Furthermore,  predictions of the model for so far unmeasured open charm mesons, baryons and nuclei in AA collisions at $\sqrt s = 5.02$ TeV and different centralities have been calculated in \cite{Andronic:2021erx}.  
{Alternative theoretical models for heavy-flavour mesons and baryons are based on the concepts of quark coalescence and recombination \cite{Plumari:2017ntm,He:2020umv,He:2019vgs}. }

Given recent data on open charm hadron production yields in different colliding systems from pp, pA to AA and for a broad energy range we analyse and identify, in the present paper, a newly emerging systematics of open charm production. Although contained in the original formulation of SHMc, this systematics was, until now, well hidden in the highly non-linear charm-balance equation ~\cite{Braun-Munzinger:2000eyl,Andronic:2003zv}
\footnote{This applies particularly for more peripheral collisions or lower energies, where the canonical formulation of thermodynamics is applicable.}. We further show that, although the SHMc has been originally proposed to describe heavy flavor production in high-energy nuclear collisions, it can be 
extended, and successfully applied to open charm production in small systems including pp collisions.

To this end we derive a direct link between the rapidity density of open charm hadron yields, $dN_i/dy$, carrying charm quantum-number $|c|=1$  and the rapidity density of charm-anticharm quark pairs, $dN_{c\bar c}/dy$ produced in the initial hard parton scatterings.
Dropping (percent level) higher order terms in pp, pA and AA collisions, $dN_i/dy$ then scales, in leading order,  with $dN_{c\bar c}/d\eta$ and the slope is quantified by the appropriate thermal density ratio calculated at the QCD chiral crossover temperature.
Furthermore,  we show that in high energy collisions and within uncertainties, the rapidity density $dN_i/dy$  of open charm hadrons exhibits the power-law scaling with charged-particle pseudo-rapidity density. Interesting properties and regularities emerge from presently available data on different ratios of open charm rapidity densities in high-energy collisions: such ratios are independent of collision energy and system size, in good agreement with predictions from the SHMc. We discuss observed quantitative deviations between data and SHMc predictions based on the 
Particle Data Group (PDG) ~\cite{PDG2024}  input of the charm hadron mass spectrum. 

 The paper is organized as follows: In the next Section, we introduce the SHMc formalism including the charm balance equation and its approximate analytic solution. Based on this new approach a connection is obtained between the rapidity density of charmed hadrons and the multiplicity of charged-particles produced in the collision.
 In Section \ref{sec:M-dependence} the open charm multiplicity ratios in high-energy collisions are discussed and compared with data. In
 Section \ref{sec:yields} we formulate the open charm production yield scaling relations. In Section \ref{summary} we summarize our results and provide conclusions and an outlook.

\section{Statistical Hadronization Model of charm production}
\label{sec:SHM_hq}
Our starting point is the charm balance equation, originally derived in ~\cite{BraunMunzinger:2000px,Braun-Munzinger:2000eyl} and further developed in ~\cite{Andronic:2021erx}:
\begin{equation}
  \begin{aligned}
    N_{\ccBar} = \frac{1}{2}& g_c V \sum_{h_{oc,1}^i} n^{{\rm th}}_i \,
    + \, \frac{1}{2} g_c^2 V \sum_{h_{oc,2}^k} n^{{\rm th}}_k + \, g_c^2 V \sum_{h_{hc}^j} n^{{\rm th}}_j \,,
  \end{aligned}
  \label{eq:balance}
\end{equation}
where $N_{\ccBar}\equiv \ud N_{\ccBar}/\ud y$ denotes the rapidity density of charm 
quark pairs produced in early, hard collisions and the (grand-canonical) thermal densities from the SHMc for open and hidden charm hadrons are given by $n_{i,j,k}^{{\rm th}}$.
The factor $g_c$ is the off-chemical equilibrium fugacity introduced to guarantee that the final number of charm quark-antiquark pairs bound in the produced hadrons is the same as $N_{c\bar c}$. The parameter $V$ is the hadronisation volume of one unit of rapidity of the fireball. The summation runs over all: open charm states $h_{oc,1}^i = D, D_s, \Lambda_c, \Xi_c, \cdots, \bar{\Omega}_c$ with one valence charm or anti-charm quark, over hidden charm states $h_{hc}^j = J/\psi, \chi_c, \psi',\cdots$, and over open charm states $h_{oc,2}^k = \Xi_{cc} \cdots, \bar{\Omega}_{cc}$ with two charm or anti-charm quarks. States with 3 charm or anti-charm quarks can be treated in complete analogy ~\cite{Andronic:2021erx} but we do not discuss those here as their contribution to the sum is negligible.

The charm balance equation should generally contain canonical corrections whenever the number of charm pairs is not large compared to unity~\cite{Braun-Munzinger:2000eyl,Gorenstein:2000ck,Braun-Munzinger:2003pwq}.
Then, Eq.~\ref{eq:balance} needs to be modified accordingly. To that end we define
\begin{equation}
\begin{aligned}
 N_{oc,\alpha} =V\,g_c^\alpha \,\sum_{h_{oc,\alpha }^i} n^{{\rm th}}_i,\,~~~
 N_{hc} =  V\,g_c^2 \sum_{h_{hc}^j} n^{{\rm th}}_j,
 \label{eq:charm_numbers}
 \end{aligned}
\end{equation}
with $\alpha=\{1,2\}$. The quantity $N_{oc,\alpha}$ is the thermal rapidity density of all charm quarks bound in hadrons $h_{oc,\alpha}^i$ with  $\alpha$  charm or anti-charm quarks, and $N_{hc}$ is the thermal rapidity density of charm-anticharm quark pairs bound in hidden charm hadrons $h_{hc}^j$.
Then, the modified charm balance equation using the canonical corrections reads:
\begin{equation}
N_{c\bar{c}}= \frac{1}{2} \sum_{\alpha = 1,2} N_{oc,\alpha} \frac{I_\alpha(N_{oc,1})} {I_0(N_{oc,1})} \, + \, N_{hc},
\label{eq:canonical}
\end{equation}
where we have assumed that the thermal density of particles and anti-particles contributing to the sum in Eq. \ref{eq:charm_numbers} are equal \cite{Braun-Munzinger:2003pwq}.   
Here,  $I_\alpha(x)$ is the modified Bessel function.

Solving Eq.~\ref{eq:canonical} for $g_c$ then, determines the charm fugacity factor 
for a given temperature and volume of a thermal fireball.   The rapidity density of open charm hadrons of type $ h_{oc,\alpha}^i $ with $\alpha=1,2$ charm quarks can then be obtained from the computed thermal densities $n_{i}^{\rm th}$ as:
\begin{equation}
   \frac{\ud N(h_{oc,\alpha}^i)}{\ud y} =g_c^\alpha \, V \, n^{{\rm th}}_i \frac{I_{\alpha}(N_{oc,1})}{I_0(N_{oc,1})}.
\label{eq:yieldsoc}
\end{equation}
For hidden charm states Eq.~\ref{eq:yieldsoc} reduces to
\begin{equation}
    \frac{\ud N(h_{hc}^j)}{\ud y} = g_c^2 \, V \, n^{{\rm th}}_j.
\label{eq:yieldshc}
\end{equation}
The essential difference between the SHMc and conventional thermal particle production is due to the fugacity factor $g_c$,  which through Eq.~\ref{eq:balance} guarantees conservation of the number of $c\bar c$ pairs from the initial partonic to the final hadronic state.

A detailed analysis of different experimental conditions of heavy ion collisions has shown that $g_c$ implies large enhancements of charm hadron yields compared to what is obtained in a purely thermal case. For central Pb-Pb collisions at LHC energies, the magnitude of $g_c$ is larger than 30. This predicted enhancement
was shown to describe very well data in heavy ion collisions at different energies and collision centrality \cite{Andronic:2003zv,BraunMunzinger:2000dv,Andronic:2011yq,Andronic:2017pug}.  

The factor $g_c$, quantifying yields in Eqs.~\ref{eq:yieldsoc} and \ref{eq:yieldshc}, is obtained by solving a 
non-linear balance equation Eq.~\ref{eq:canonical}. This requires experimental input for the initial number of charm quark-antiquark pairs and values of thermal parameters at chemical freeze-out, which are linked to the corresponding collision energy and colliding systems.     

From recent SHMc analysis of different charm hadron production yields in central Pb-Pb collisions at $\sqrt{s_{NN}}= 5.02$ TeV \cite{Andronic:2021erx}, one can conclude that the sum of charm-two and hidden-charm contributions in the balance equation is small, not exceeding 3$\%$. For the present investigation, where we aim to get analytic results,  we will neglect all terms with $\alpha>1$.  Then, Eq.~\ref{eq:canonical} can be solved for $g_c$ with only  the $\alpha=1$ term, giving that: 

\begin{equation}
g_c \, V \, \frac{I_{1}(N_{oc,1})}{I_0(N_{oc,1})}\simeq
\frac{2N_{c\bar{c}}}{n_{oc,1}^{\rm tot}},
\label{eq:gc}
\end{equation}
where 
\begin{equation}
  \begin{aligned}
    n_{oc,1}^{\rm tot} = &\sum_{h^i_{oc,1}} n^{{\rm th}}_i \,
      \end{aligned}
  \label{eq:tot}
\end{equation}
is the total thermal density of all particles carrying charm quantum-number,  $|c|=1$. From  Eqs.~\ref{eq:gc} and \ref{eq:yieldsoc}, the rapidity density of open charm hadrons with $|c|=\pm 1$ is obtained as:
\begin{equation}
   \frac{\ud N(h_{oc,1}^i)}{\ud y} \simeq 2\frac{n^{{\rm th}}_i}{ n_{oc,1}^{\rm tot}}N_{c\bar{c}},
\label{eq:main}
\end{equation}
thus, in leading order, it is independent of the volume of the fireball and canonical corrections related to exact charm conservation. The rapidity density of all open charm particles then scales with the number of $c\bar c$ pairs produced in the initial hard scatterings. The proportionality factor is fully calculated in the SHMc with thermal parameters specific for chemical freeze-out conditions at the corresponding  (high) energy.

In the following, we will discuss the production yield systematics of $D^0, D^+, D^{*+}$ and $D^+_s$ mesons, as well as $\Lambda^+_c$ baryons in pp, pPb and Pb-Pb collisions at different LHC energies and in pp and Au-Au collisions at $\sqrt{s_{NN}} = 200$ GeV from RHIC. 
This will test the new notion that the SHMc framework with canonical thermodynamics is appropriate to also describe charmed hadron production in high energy pp and pPb collisions.

\section{SHMc and open charm multiplicity ratios in high-energy collisions}
\label{sec:M-dependence}

The thermal part in Eq. \ref{eq:main},  calculated in the grand canonical ensemble,
depends in general on temperature and the value of the chemical potential $\vec \mu=(\mu_B,\mu_S,\mu_Q)$ linked to the conservation of baryon number, strangeness and electric charge. At high energy, however, within a few units around mid-rapidity data are completely dominated by hadrons with valence quarks produced in the collision. 
As a consequence, at the LHC energies $\vec \mu\simeq 0$,  and the only thermal parameter characterising the density ratios in  Eq. \ref{eq:main} is the temperature. 

{From a detailed analysis of light flavor particle yields at the LHC it is well established that they are frozen in at the QCD phase boundary \cite{Andronic:2017pug,Cleymans:2020fsc}. Thus, the temperature parameter in Eq. \ref{eq:main} is expected to coincide with the chiral crossover temperature $T_c=156.5$ MeV calculated in first principle LQCD \cite{HotQCD:2018pds}. 
Furthermore, in heavy ion collisions, 
the chemical freeze-out temperature for  ${\sqrt {s_{NN}}} \geq 10$ GeV is essentially energy independent \cite{Andronic:2017pug,Andronic:2021dkw}.  At
LHC energies, this temperature is also common for all colliding systems \cite{Cleymans:2020fsc}. }

Because of the above, and from Eqs. \ref{eq:main} and \ref{eq:yieldsoc} 
one arrives at a new and key prediction of the SHMc:  the ratios of 
rapidity densities of different particle species carrying charm quantum-number $|c|=1$ should be independent of system size and energy in high-energy collisions.

\begin{figure}
\includegraphics[scale=0.45]{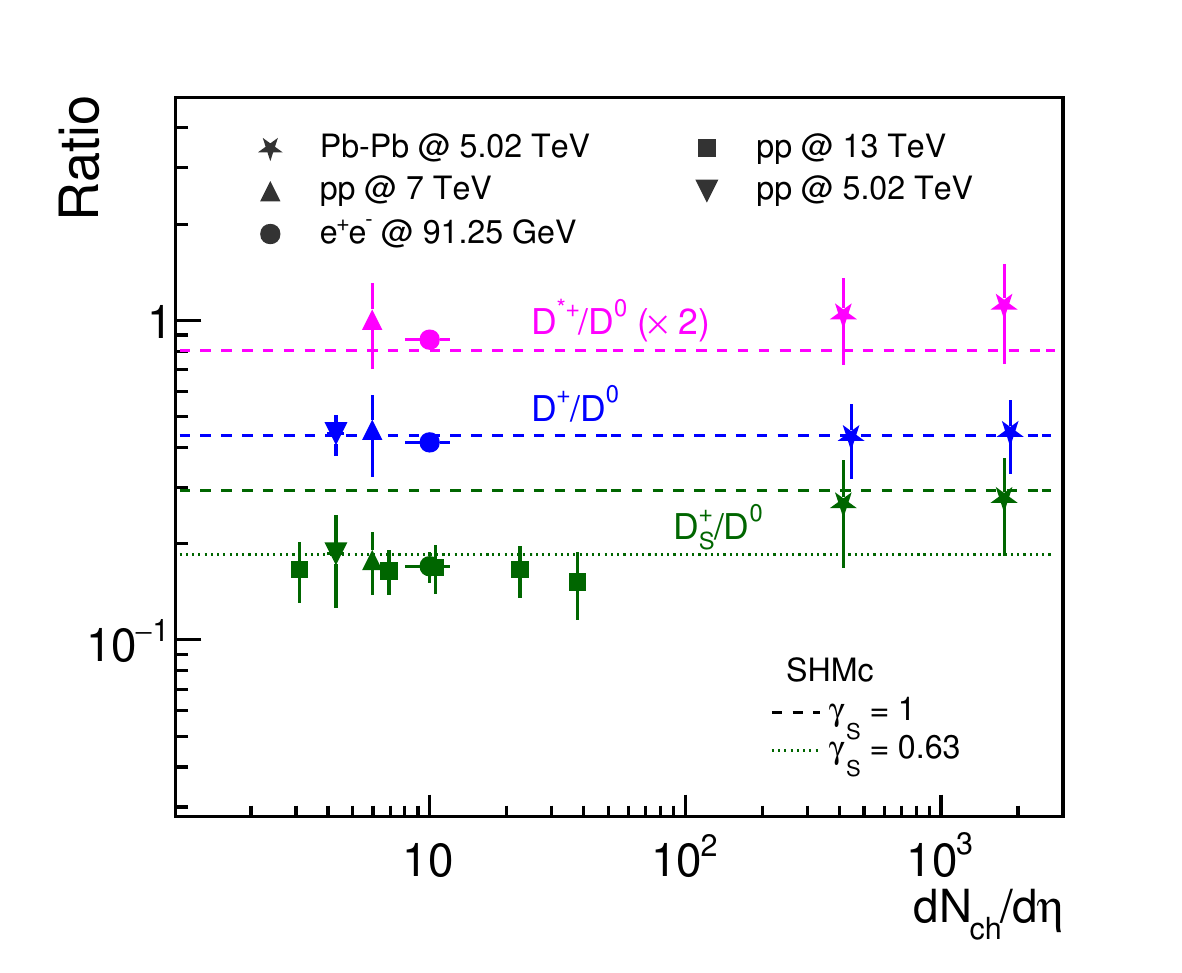}
\vskip -0.4 cm
\caption {Ratios of rapidity densities of $D^+, D^{*+}$ and $D^+_s$ to $D^0$ mesons plotted as a function of the charged-particle pseudo-rapidity density. For clarity, the ratio $D^{*+}/D^0$ has been multiplied by a factor of 2 and the points for $D^{+}/D^0$ ratio in Pb-Pb collisions have been displaced horizontally for better visibility. The $pp$ \cite{
ALICE:2019nxm,ALICE:2021mgk,ALICE:2017olh,
ALICE:2021npz,ALICE:2023sgl} and Pb-Pb \cite{ALICE:2021rxa,
ALICE:2021kfc} data are from the ALICE experiment. The $e^+e^-$ data are from the compilation of LEP data \cite{Gladilin:2014tba}. The dashed horizontal lines are SHMc predictions for a temperature $T_c=156.5$ MeV.  The dotted line is the SHMc value calculated with the strangeness undersaturation factor,  $\gamma_S=0.63$, see text. }  
\label{fig:ratioDtoD0}
\end{figure}

\begin{figure}
\includegraphics[scale=0.45]{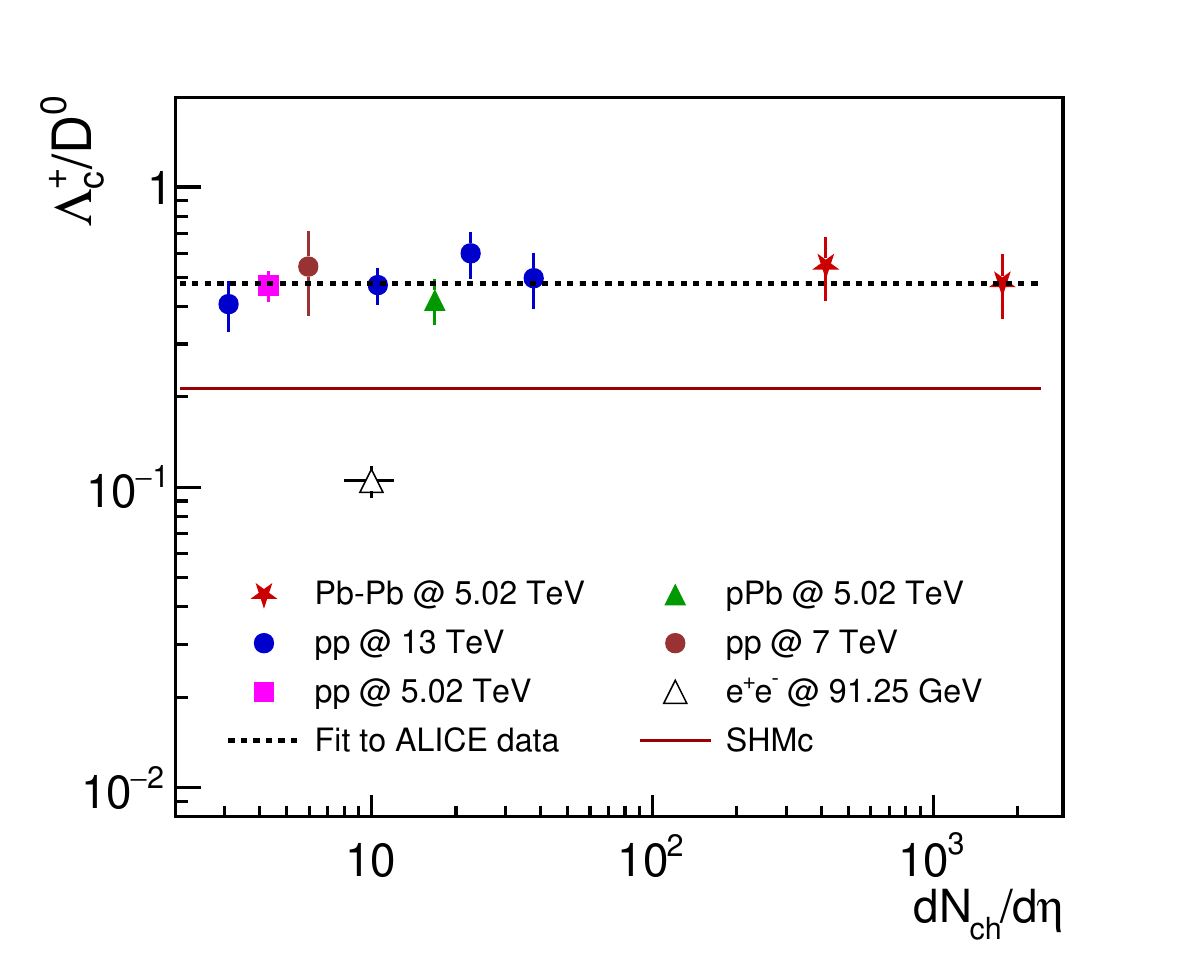}
\vskip -0.4 cm
\caption{Ratio of $\Lambda^+_c/D^0$ rapidity densities obtained by the ALICE experiment in minimum bias pPb, Pb-Pb collisions at different centralities \cite{ALICE:2016yta,ALICE:2022exq,ALICE:2017thy,
ALICE:2021bib}, and in pp collisions at various energies and multiplicities \cite{ALICE:2019nxm,ALICE:2021mgk,ALICE:2017olh,ALICE:2017thy,ALICE:2021npz,ALICE:2023sgl}. The ratios are plotted as a function of the charged-particle pseudo-rapidity density. The $e^+e^-$ value is from the compilation of LEP data \cite{Gladilin:2014tba}. The solid horizontal line is the SHMc prediction with PDG data input\cite{PDG2024} at $T_c=156.5$ MeV. The short dashed horizontal line is a fit to the data, see text. }
\label{fig:ratioLtoD0}
\end{figure}

In Figs. \ref{fig:ratioDtoD0} and \ref{fig:ratioLtoD0} we present ratios of rapidity densities of $D^+,D^{*+},D^+_s$ mesons and $\Lambda^+_c$ baryons to that of $D^0$ mesons. Data from pp collisions at different energies and multiplicities and from Pb-Pb collisions at different centralities are from the ALICE experiment \cite{ALICE:2016yta,
ALICE:2019nxm,ALICE:2021mgk,ALICE:2017olh,
ALICE:2023sgl,ALICE:2021npz,ALICE:2021rxa,ALICE:2021kfc,ALICE:2021bib,ALICE:2022exq,ALICE:2017thy}. The $e^+e^-$ data are from the compilation in \cite{Gladilin:2014tba}.
As predicted in the SHMc, these ratios, plotted as a function of the associated charged-particle pseudo-rapidity density, are for charmed mesons constant and independent of collision energy and colliding system within uncertainties. For the $\Lambda^+_c/D^0$ ratio, this is also the case for hadronic collision systems, while the value for $e^+e^-$ is lower (see discussion below).

The results in  Figs. \ref{fig:ratioDtoD0} and \ref{fig:ratioLtoD0} are quantified in the SHMc as density ratios,  $n^{th}_i/n^{th}_j$. The total density of open charm species $i$ is a sum of the prompt thermal component $n^p_i$ and the contribution of resonances decaying by strong interaction to particle $i$:
\begin{equation}
\label{eq:reso}
n^{th}_i
= n^p_i + \sum_j Br(j\rightarrow i) n^{r}_j, 
\end{equation}
where $Br(j\rightarrow i)$ is the decay branching ratio of resonance $j$ to particle $i$. We use the THERMUS package \cite{Wheaton:2004qb,Thermus} updated with recent results in the charm and beauty sector as given in the  PDG review~\cite{PDG2024}. The results agree within uncertainties with those quoted in ~\cite{Andronic:2021erx}.

Considering contributions of all open charm hadrons and resonances listed by the PDG
\cite{PDG2024} we are displaying in Figs. \ref{fig:ratioDtoD0} and \ref{fig:ratioLtoD0} the SHMc ratios of rapidity densities of $D^+,D^{*+},D^+_s$ and $\Lambda^+_c$ to $D^0$ at $T_c\simeq 156$ MeV. The SHMc results for $D^+/D^0$ and $D^*/D^0$ are in good agreement with the measured values. However, data on  $\Lambda^+_c/D^0$ are larger by a factor of $2.2\pm 0.15$, compared to the SHMc predictions. The observed charm baryon enhancement relative to the SHMc using the PDG data input for resonances was already observed in ~\cite{He:2020umv,He:2019vgs,He:2019tik} and ~\cite{Andronic:2021erx} and tentatively attributed to missing resonances. Independently, already the first LQCD results on different fluctuation observables in the charm-baryon sector have 
indicated missing resonances \cite{Bazavov:2014yba,HadronSpectrum:2012gic}. The relativistic quark model of hadrons (RQM) \cite{Ebert:2011kk}, 
as well as LQCD \cite{Padmanath:2013bla} yield a significantly larger number of charmed baryon resonances than those observed experimentally so far.
Recent LQCD results in the charm baryon sector also suggest increased charm baryon states at  $T_c$  relative to PDG \cite{Bazavov:2023xzm,Sharma:2025zhe}.
Implementing the additional charmed baryon resonances into the statistical model ~\cite{He:2020umv,He:2019vgs,He:2019tik} or increasing the statistical weights of the PDG states 
accordingly, yields an increase by a factor of approximately two in the $\Lambda^+_c$ rapidity density.

Indeed, considering the RQM for missing resonances and modelling the  branching ratios  to their ground states it was shown in  \cite{He:2019tik}, 
that
the ratio of $\Lambda_c/D^0\simeq 0.44 $ at $T=160$ MeV and   $\Lambda_c/D^0\simeq 0.57$ at $T=170$ MeV, approximately doubling their  PDG values at corresponding temperatures.
Comparing these values with the linear fit to data,  $\Lambda_c/D^0\simeq 0.48\pm 0.04$ in Fig. 2, one notices that this RQM  model is consistent with data at $160\leq T < 170$ MeV. We note, however, that in such an approach there are additional uncertainties linked to assumptions on the decay branching of resonances, and likely still not complete charmed baryon mass spectrum in the RQM. 

That is in the following,  instead of assuming the RQM mass spectrum and adjusting the temperature range to reproduce data within errors, we will follow the LQCD observation,
that at $T_c=156.5\pm 1.5$ MeV the thermodynamic pressure of charmed baryons, relative to that with the PDG input,  is larger by a factor of $1.948\pm 0.234$  due to missing resonances \cite{Sharma:2025zhe}. 
We assume that similar is also the case for $\Lambda_c$ density.  We will fix the temperature to $T_c$  and rescale the SHMc predictions for the $\Lambda_c/D^0$ ratio by a phenomenological factor of $2.2\pm 0.15$,  to effectively account for the contribution from yet unknown resonances,  and to match $\Lambda_c/D^0$ data.

With increasing accuracy of charmed hadron data and a more complete experimental knowledge of the charmed baryon mass spectrum, a more precise determination 
 of the freezeout temperature of charmed hadrons will be possible in the future. In particular, it will allow us to distinguish if they freezeout at the chiral crossover, as assumed in these studies,  or at higher temperatures as suggested e.g.  in 
\cite{He:2019tik}.


\begin{figure}
\includegraphics[scale=0.45]{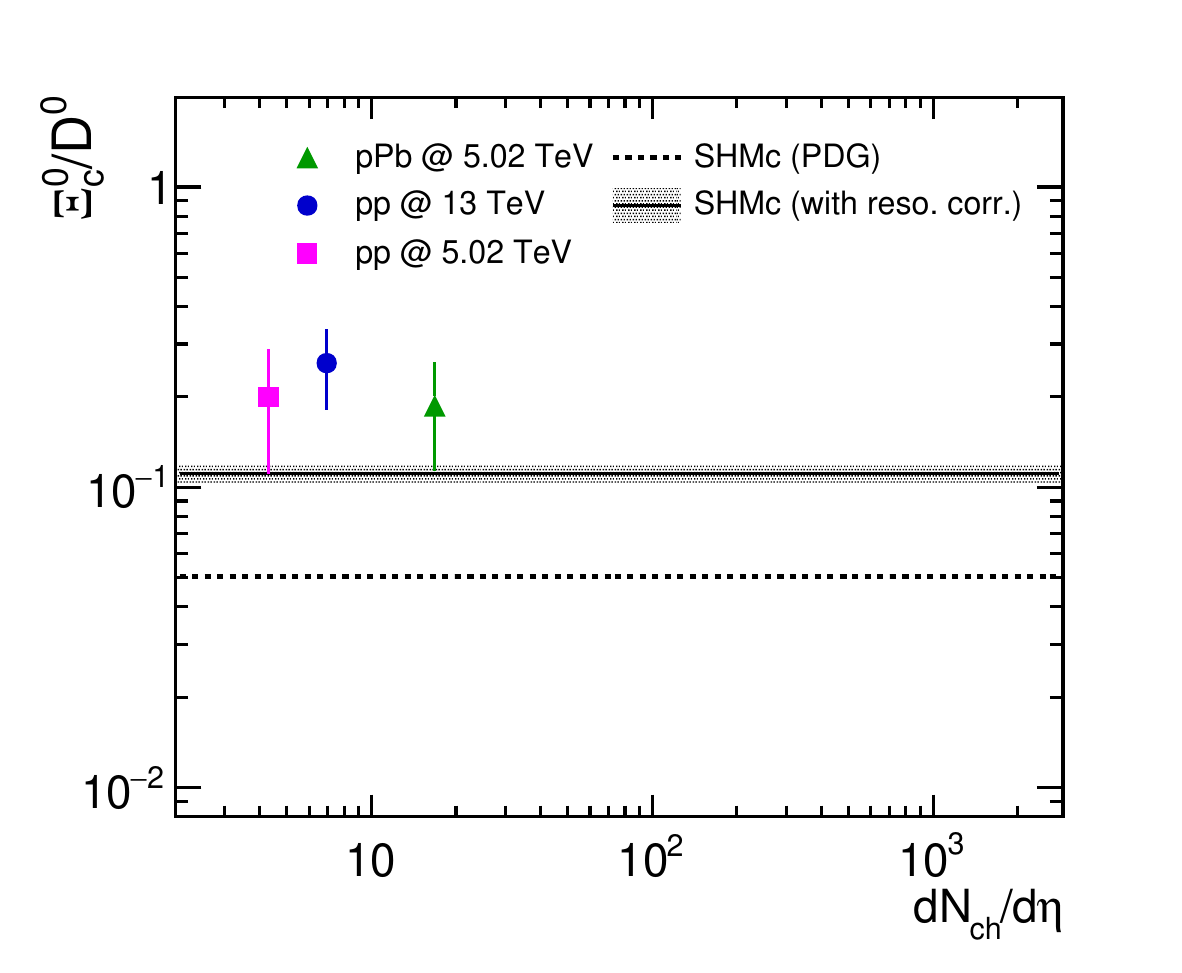}
\vskip -0.4 cm
\caption{
{Ratio of $\Xi_c^0/D^0$ rapidity densities obtained by the ALICE experiment in pp and pPb collisions at $\sqrt s=5.07$ TeV and in pp at $\sqrt 13$ TeV  \cite{ALICE:2024ocs,ALICE:2023sgl}. 
The ratios are plotted as a function of the corresponding charged-particle pseudo-rapidity density.  The dotted-horizontal line is the SHMc prediction at $T_c=156.5$ MeV with the  PDG input.  The shaded band is the SHMc result with missing charmed-baryon resonances included, see text.
 }}
\label{fig:3n}
\end{figure}

{
Such an increase of open charm baryon yields beyond the SHMc predictions with the PDG input is also there in the $\Xi_c^0/D^0$ ratio. 
In Fig. ~\ref{fig:3n} we show $\Xi_c^0/D^0$ rapidity density ratio at midrapidity in pp and pPb collisions at $\sqrt s =5.02$ TeV and in pp collisions at $\sqrt s =13$ TeV obtained within ALICE experiment \cite{ALICE:2024ocs,ALICE:2023sgl}. 
The ratio is compared with the SHMc predictions with the PDG input and after rescaling the  $\Xi_c^0$-density by the same factor as found for  $\Lambda_c^+$  in Fig. 2 to account effectively for missing resonances. 
In the last  case, the pp and pPb data at  $\sqrt s =5.02$ TeV are  consistent with the SHMc results within 1$\sigma$, however, there is a more than 2$\sigma$ deviation from the pp value at  
$\sqrt s =13$ TeV. With new results from Run 3  of the LHC,  one expects data for  $\Xi_c$ yields with reduced uncertainties to allow for a more conclusive interpretation of open charm baryon production and their hadronization.  
}


Even more surprising, however, is an observation in Fig.~\ref{fig:ratioDtoD0} that data in pp collisions for $D^+_s/D^0$ are suppressed by a factor {$\gamma_s \simeq 0.63\pm 0.1 $}   
relative to the SHMc value calculated at $T_c$. Furthermore,  {Pb-Pb data for central and semi-central collisions are consistent with the SHMc with $\gamma_s=1$.}  
This is unexpected since in the (u,d,s) sector strangeness production in pp and Pb-Pb collisions at the LHC is consistent with chemical equilibrium production i.e. with $\gamma_s=1$, see ~\cite{Andronic:2005yp,Andronic:2017pug}. The observed small suppression of single strange to non-strange meson ratios from central AA to pp collisions and its $dN_{ch}/d\eta$ scaling was quantified by accounting for exact strangeness conservation \cite{Cleymans:2020fsc} in canonical thermodynamics and amounts for minimum bias pp collisions for singly strange hadrons only {up} to  10\% reduction. 
On the other hand, a suppression by a similar factor, i.e. {$\gamma_s\simeq 0.66$}, was observed in thermal analysis of $e^+e^-$ scattering data \cite{Becattini:2008tx,Andronic:2009sv,Redlich:2009xx} for singly strange un-charmed and charmed hadrons.

{For heavy ion  collisions, {only a limited set of} fully $p_T$-integrated data exists for the $D^+_s/D^0$ yield ratio \cite{ALICE:2021kfc}. Furthermore, }
an enhancement relative to pp collisions was observed by the ALICE experiment when comparing  $D^+_s/D^0$ ratios at fixed $p_T$ for $p_T>3.5 $ GeV/c in central Pb-Pb collisions at $\sqrt {s_{NN}} = 5.02$ TeV and scaled pp collisions at $\sqrt {s_{NN}}=7$ TeV \cite{ALICE:2018lyv}. For this data set, good agreement in the measured $p_T$ range with SHMc predictions was reported in ~\cite{Andronic:2021erx}. Recently the ratio $D^+_s/D^0$ was measured for central Pb-Pb collisions down to $p_T=2$ GeV/c, and the yield was extrapolated to $p_T$ = 0 using model shapes of the spectrum~\cite{ALICE:2021kfc}. As currently 70 \% of the yield is measured, the extrapolation contributes a 24 \% uncertainty to the extrapolated total rapidity density. However, this extrapolated rapidity density is in good agreement with the SHMc prediction, with the central value of data amounting to 85 \% of the SHMc value.   
The STAR Collaboration has reported on the measurement of this ratio in Au-Au collisions at $\sqrt {s_{NN}} =0.2$ TeV for yields integrated within a window of $ 1.5\le p_T\le 5.0$ GeV/c,  and for different centrality classes \cite{STAR:2021tte}. A fit of the $D^+_s/D^0$ yield ratio in Au-Au collisions to a constant value given by the STAR Collaboration is by a factor of two larger than the pp values shown in Fig.~\ref{fig:ratioDtoD0}.

The LHCb Collaboration recently measured the multiplicity dependence of the ratio $D^+_s/D^+$ for pPb collisions at forward rapidity ~\cite{LHCb:2023rpm}. The data exhibit a strong rise with multiplicity in the transverse momentum range $2.0 < p_{\rm T} < 4$ GeV/c and $6.0 < p_{\rm T} < 8$ GeV/c and for $dN_{ch}/d\eta$ in the range between 30 and 60. 
Once  more complete data on $p_{\rm T}$ integrated ratios for systems including the $D^+_s$ meson are available, 
it will be important to investigate in detail the role of the $D^+_s$ in charm hadron production systematics.

The $\Lambda^+_c/D^0$ ratio in $e^+e^-$ scattering shown in Fig. \ref{fig:ratioLtoD0} is lower than pp and AA data.
This may indicate that the population of additional charm baryon states beyond PDG is suppressed in  $e^+e^-$ collisions. Furthermore, the $D^+_s/D^0$ yield ratio in $e^+e^-$  coincides with pp value and hence, as mentioned above, is suppressed by {$\gamma_s \simeq 0.63$} relative to SHMc predictions. We note, that hadron yield data in $e^+e^-$, including charm and bottom production, were quantified by the Hadron Resonance Gas Model formulated in the canonical ensemble with exact conservation of all five additive quantum-numbers, and with the $\gamma_s<1$ suppression factor \cite{Becattini:2008tx,Andronic:2009sv,Redlich:2009xx,Andronic:2008ev,Becattini:2010sk}.


{In the following,  
we will compare the model predictions and data for open charm production yields and establish their scalings.
The results will be shown with the PDG resonance input in the charmed-baryon sector and compared with 
the adjusted mass spectrum to account for missing charmed-baryon resonances as identified in Fig.~\ref{fig:ratioLtoD0}.}

\section{Open charm production yields}
\label{sec:yields}

In high-energy collisions, the thermal densities in Eq.
\ref{eq:main}
depend only on $T\simeq T_c$.  Thus, to quantify the rapidity densities of open charm hadrons, following Eq. \ref{eq:main}, one needs to specify the rapidity density of charm quark-antiquark pairs, $N_{\ccBar}$. 
In AA (as in pp) collisions $N_{\ccBar}$ is a quantity that should be determined by measurement of all hadrons with open or hidden charm. Such experimentally determined $N_{\ccBar}$ values would already include all nuclear effects in charm production as compared to pp collisions, and also account for possible additions to the charm yield from thermal \ccBar~ production in the QGP, as well as, potential losses due to charm quark annihilation. The latter is, however, expected to be negligible at the sub percent level ~\cite{Andronic:2007bi}.
In practice, using this prescription is however difficult, since the measurement of all open and hidden charm hadrons should be performed without cuts in transverse momentum, to keep systematic uncertainties due to extrapolations small. While this has been accomplished for pp collisions in ALICE~\cite{ALICE:2023sgl} with a 10\% precision (to which remaining necessary extrapolations contribute only 2\%), achieving such precision measurements of $N_{\ccBar}$ for Pb-Pb collisions down to $p_T$ = 0 is one of the priorities for the upgraded ALICE experiment. 
\begin{figure}
\centering
\includegraphics[scale=0.45]{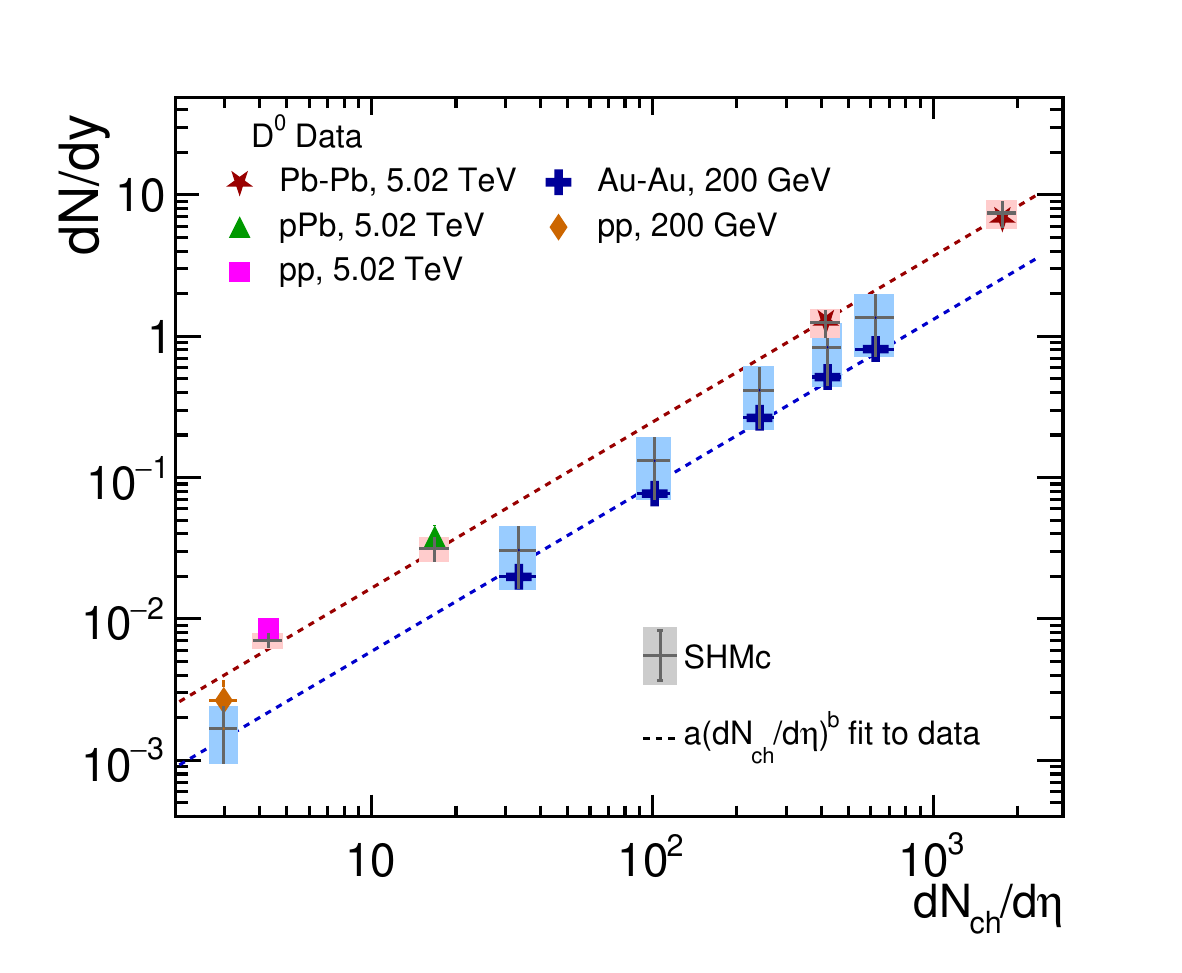} 
\vskip -0.4 cm
\caption{Rapidity density for $D^0$ mesons at mid-rapidity for different collision systems, plotted at the corresponding 
 the pseudo-rapidity density of charged-particles. The Pb-Pb, minimum-bias pp and pPb data at $\sqrt {s_{NN}}=5.02 $ TeV are from the ALICE experiment \cite{ALICE:2021rxa,
ALICE:2019nxm,
ALICE:2021mgk,ALICE:2017olh, Acharya:2019mno}.
The densities of minimum-bias pp and Au-Au collisions at different centralities at $\sqrt {s_{NN}}=200$ GeV are obtained from the  STAR data \cite{STAR:2018zdy,STAR:2014wif}, as described in the text and listed in Table \ref{table:D0AuAu}. The SHMc results at $T_c=156.5$ MeV are shown as horizontal lines inside boxes 
with their height corresponding to model uncertainties. 
 The short dashed lines represent power-law fits to data, see text.
}
\label{fig:ratioLtoD03}
\end{figure}

In the absence of a measured charm production cross-section in AA  collisions, based on the notion that charm quarks are produced in early hard collisions, we apply the concept of scaling with the number of binary collisions for a given collision geometry. We obtain $N_{\ccBar}$ at mid-rapidity from the measured minimum-bias charm cross-section $ \sigma^{pp}_{c\bar{c}}\equiv \langle\der \sigma_{c\bar{c}}/\der y\rangle$ 
at mid-rapidity in pp collisions, multiplying it with the appropriate nuclear thickness function for a given centrality interval, $T_{AA}$. Furthermore, $\sigma^{pp}_{c\bar{c}}$ is folded with a factor $\alpha_A$ accounting for nuclear modification effects, such as shadowing, energy loss or saturation effects, typically obtained from pA collisions. Thus, in heavy ion collisions,  
$N_{c\bar c}^{AA}=\alpha_A \, \sigma^{pp}_{c\bar{c}} \,T_{AA}$. 

Assuming that charm hadronisation in pp collisions also follows thermal/statistical concepts, one can use Eq. \ref{eq:main} to calculate $dN_i^{pp}/dy$ of open charm hadrons, with the according $N_{c\bar c}^{pp}=\sigma^{pp}_{c\bar{c}}/{\sigma^{pp}_{inel}}$. 

In such an extended SHMc, the rapidity density of open charm hadron species $i$  with charm $|c|=1$  in high energy pp and AA collisions can be  calculated from: 
\begin{equation}
   \frac{\ud N_i}{\ud y} =2\frac{n^{{\rm th}}_i}{ n_{oc,1}^{\rm tot}}\, N_{c\bar c},
\label{eq:mainAA}
\end{equation}
with the rapidity density of the number of $N_{c\bar c}$ pairs obtained from:

\begin{equation}
 N_{c\bar c}=\begin{cases}
 {\sigma^{pp}_{c\bar{c}}}/{\sigma^{pp}_{inel}} & \text{ in~pp } \\
\alpha_A\, \sigma^{pp}_{c\bar{c}}\,T_{AA}  ~
    &~\text{in AA}
\end{cases}
\label{eq:final}
\end{equation}
With the experimental input for the charm and inelastic cross-sections and thickness functions calculated from the Glauber model \cite{ALICE:2018tvk}, Eqs. \ref{eq:final} and \ref{eq:mainAA} constitute the SHMc prescription of rapidity densities of different open charm hadron species in pp and AA collisions. 

In Fig.~\ref{fig:ratioLtoD03} we show the $D^0$ rapidity density in minimum-bias pp and Pb-Pb, as well as pPb collisions at $\sqrt{s_{NN}}=5.02$ TeV as a function of the charged-particle multiplicity density, $dN_{ch}/d\eta$. Data from the ALICE Collaboration \cite{ALICE:2021rxa,ALICE:2019nxm,ALICE:2021mgk,ALICE:2016yta} are compared with SHMc predictions. 
The experimental input to Eqs. \ref{eq:final} and \ref{eq:mainAA} in pp and Pb-Pb collisions at $\sqrt{s_{NN}}=5.02$ TeV, i.e. the minimum-bias $\sigma_{c\bar c}^{pp}$ and $\sigma_{inel}^{pp}$ are also from the ALICE experiment  \cite{ALICE:2016yta,ALICE:2019nxm,ALICE:2021mgk,ALICE:2017olh,ALICE:2017thy,ALICE:2021rxa,ALICE:2021bib,ALICE:2023sgl,ALICE:2021dhb,ALICE:2021npz,ALICE:2012fjm}. 
For the mid-rapidity reduction factor $\alpha_A$, we have used the value of $0.65\pm 0.12$ as explained in \cite{Andronic:2021erx}.   
The thickness function $T_{PbPb}$ for different centrality classes corresponding to $dN_{ch}/d\eta$ in Fig.~\ref{fig:ratioLtoD03} can be found in Ref. \cite{ALICE:2021rxa}. 

The $D^0$ yield in pPb can be obtained in the SHMc by calculating $N_{c\bar c}$ from the measured ratio of ${\sigma^{pPb}_{c\bar{c}}}/{\sigma^{pPb}_{inel}}$ in pPb at the corresponding $dN_{ch}/d\eta$~\cite{ALICE:2012xs}. Indeed, taking ${\sigma^{pPb}_{c\bar{c}}}=151\pm 26$ [mb]
 and ${\sigma^{pPb}_{inel}}=2.10\pm 0.055$ [b],  from  the ALICE experiment \cite{ALICE:2016yta,ALICE:2018tvk},
one gets: $dN_{D^0}/dy\simeq 0.035$. 
{The experimental value from the ALICE Collaboration for the rapidity density of $D^0$ in pPb collisions was found as: 
$dN_{D^0}/dy= 0.038 \pm 0.006$  \cite{ALICE:2016yta,Acharya:2019mno}. }
\begin{table}
\vskip 0.3cm
\centering
\begin{tabular}{|l|r|r|r}
\hline
System & Centrality & $dN_{D^0}/dy$ \\
\hline
$pp$ & $MB$ & $(2.65\pm 1.02)\times 10^{-3}$ \\
\hline
\multirow{5}{*}{Au-Au} & $0-10\%$ & $0.811\pm 0.121$ \\
& $10-20\%$ & $0.514\pm 0.078$ \\
& $20-40\%$ & $0.265\pm 0.037$ \\
& $40-60\%$ & $0.077\pm 0.011$ \\
& $60-80\%$ & $0.020\pm 0.0034$ \\
\hline
\end{tabular}
\caption{\label{tab:widgets} Rapidity densities, $dN_{D^0}/dy$ obtained after integrating the $p_T$ differential invariant distributions in pp and Au-Au collisions at $\sqrt{s} = 200$ GeV by the STAR Collaboration~\cite{STAR:2014wif,STAR:2018zdy}.
} 
\label{table:D0AuAu}
\end{table}

In Fig.~\ref{fig:ratioLtoD03} we also show the $D^0$ rapidity density in Au-Au and minimum-bias pp data at $\sqrt{s_{NN}}=0.2$ TeV \cite{STAR:2014wif,STAR:2018zdy}. The charged-particle rapidity densities for pp and different centralities Au-Au collisions are taken from \cite{STAR:2008med}. To get $D^0$ rapidity densities at the top RHIC energy and for different collision centralities, we have fitted the $p_T$ distributions of $D^0$ measured by the STAR Collaboration with a Levy-Tsallis function~\cite{Tsallis:1987eu} and integrated them. 
\begin{figure}
\centering
\includegraphics[scale=0.45]{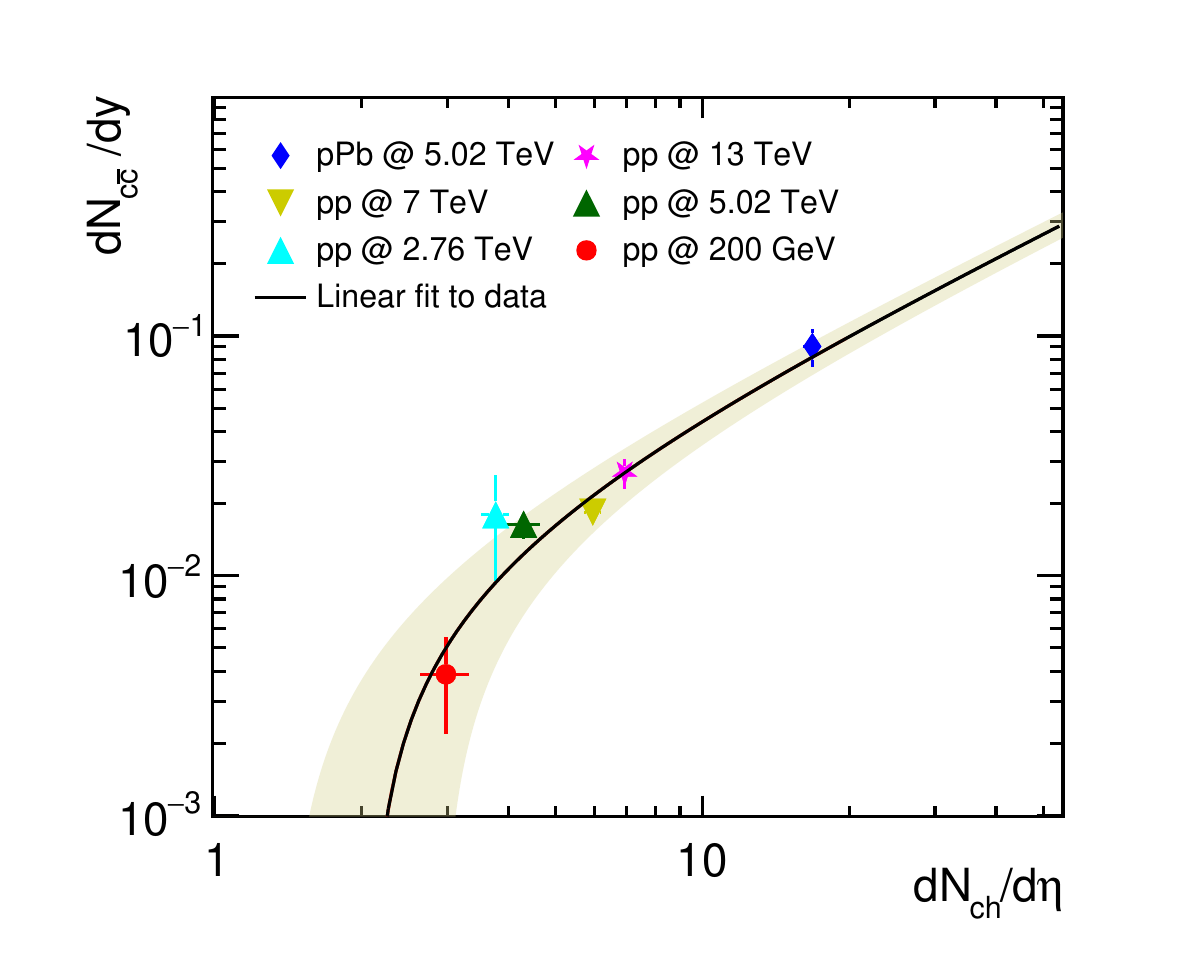}
\caption{Measured charm-anticharm rapidity density at mid-rapidity obtained from ratios of open charm, $d\sigma_{c\bar c}/dy$  and inelastic, $d\sigma_{inel}/dy$     cross-sections in pp and pPb collisions plotted at the corresponding charged-particle pseudo-rapidity density. The pPb value was normalized by the nuclear modification factor, $\sqrt \alpha_A$ to make it comparable to pp values. Data at the LHC are from the ALICE Collaboration \cite{ALICE:2016yta,ALICE:2019nxm,ALICE:2021mgk,ALICE:2017olh,ALICE:2017thy,ALICE:2021rxa,ALICE:2021bib,ALICE:2018tvk} with  $d\sigma^{pp}_{inel}/dy$   at $\sqrt {s_{NN}}=13$ TeV from the LHCb Collaboration \cite{LHCb:2018ehw}. The  pp data at $\sqrt {s_{NN}}=200$ GeV are from the STAR experiment \cite{STAR:2012nbd,STAR:2020phn}. The line is a linear fit restricted to   $3<dN_{ch}/d\eta<18$ window with the corresponding 1$\sigma$ uncertainty band.}
\label{fig:ccpp}
\end{figure}
The resulting rapidity densities of $D^0$ at RHIC are summarized in Table~\ref{table:D0AuAu}.

For the model comparison with RHIC results, the experimental inputs for Eqs. \ref{eq:final} and \ref{eq:mainAA} in minimum-bias pp  collisions at $\sqrt{s_{NN}}=0.2$ TeV are from the STAR experiment \cite{STAR:2012nbd,STAR:2020phn}. The thickness function is calculated using the Monte Carlo Glauber model {from Ref. \cite{STAR:2018zdy}}. 
{The nuclear modification factor at RHIC was extracted
from a nuclear gluon distribution function including LHC open heavy flavor data in the fit \cite{Duwentaster:2022kpv} as $\alpha_A=0.86\pm 0.15$.} 

The SHMc model predictions shown in  Fig.~\ref{fig:ratioLtoD03} are well consistent, within errors,  with Pb-Pb, pPb and pp minimum-bias data at the LHC, as well as with the corresponding Au-Au and pp data at RHIC. { The SHMc results are depicted as boxes with their height corresponding to model uncertainties linked to experimental inputs to Eqs.  \ref{eq:mainAA} and  \ref{eq:final}.

Furthermore, an interesting and for us quite unexpected feature  of  the $D^0$ rapidity density shown  in  Fig.~\ref{fig:ratioLtoD03} is 
its approximate power-law scaling, $dN/dy=a(dN_{ch}/d\eta)^b$ with the charged-particle density, $dN_{ch}/d \eta$ and power-law constants $a$ and $b$. 
Fitting to LHC data one obtains: $a=(1.1\pm 0.1)\times10^{-3}$ and $b=1.2\pm 0.02$. The data at RHIC exhibit a similar slope as found for LHC data. Fixing the mean value,  $b= 1.2$, one gets at RHIC,   $a\simeq (3.8\pm0.31)\times10^{-4}$.  Considering, however,  the rather poor precision,  for  RHIC energy, of the open charm cross-section which dominates the uncertainties in the SHMc predictions, it cannot be excluded that the power coefficients in this scaling may exhibit some $\sqrt { s_{NN}}$ dependence.
The SHMc results shown in Fig.~\ref{fig:ratioLtoD03} point in this direction. 
In the text below we will explore the observed scaling with more data for different particle species to verify the  SHMc expectation.
 }   
\begin{figure}
\includegraphics[scale=0.45]{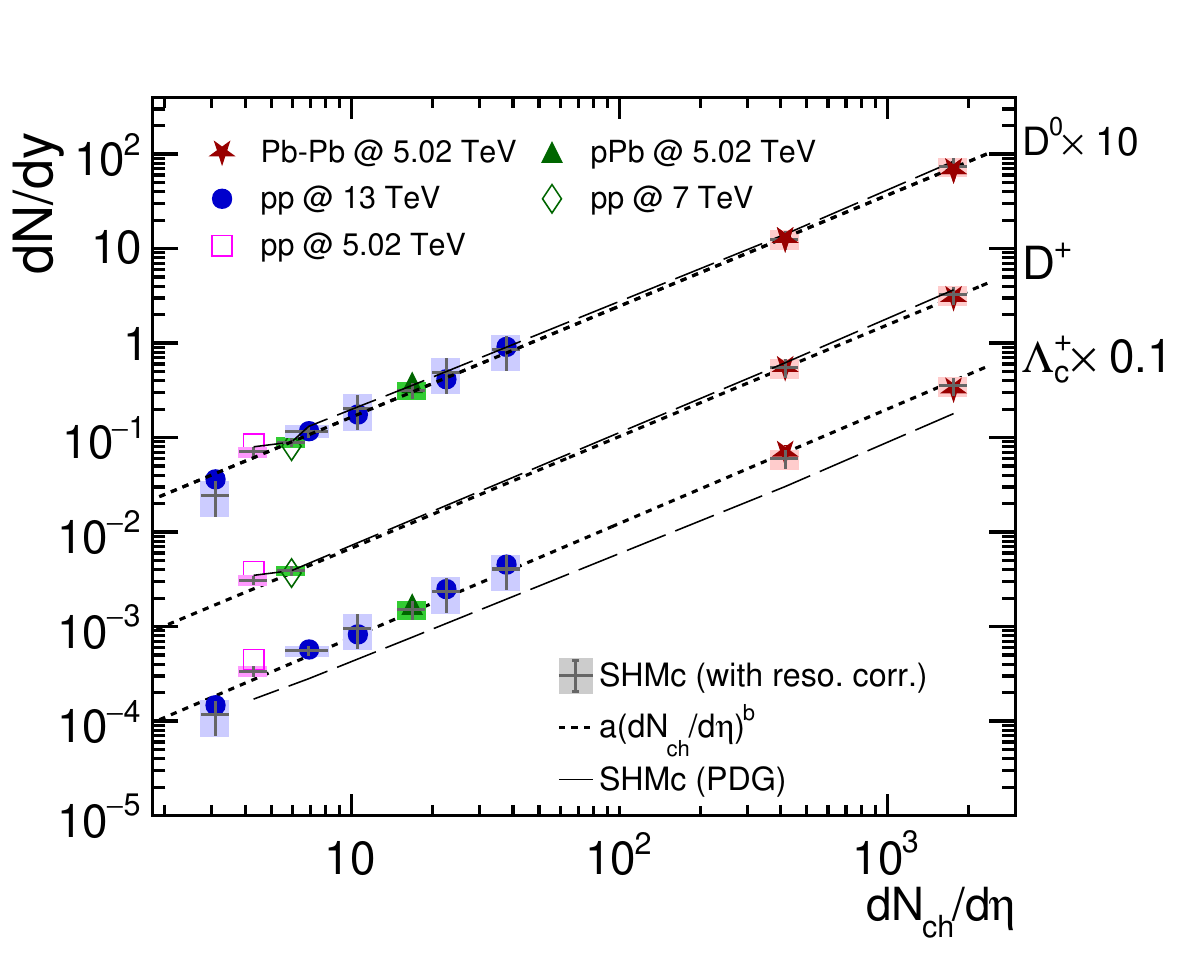}
\vskip -0.4 cm
\caption{As in Fig. \ref{fig:ratioLtoD03} including  $D^+$ and $\Lambda^+_c$ yield densities,  and pp data at $\sqrt {s_{NN}}= 7.0$ and 13 TeV for different charged-particle rapidity densities. Also shown are SHMc predictions 
{including missing charmed-baryon resonances and SHMc with  PDG input (long-dashed lines)},   see text. 
The dotted-lines are power-law fits, see text.  Data are from the ALICE experiment \cite{ALICE:2016yta,ALICE:2019nxm,ALICE:2021mgk,ALICE:2017olh,ALICE:2017thy,ALICE:2021rxa,ALICE:2021bib}. }
\label{fig:gc-scaling}
\end{figure}

The production of charm in pp collisions has been measured by the ALICE Collaboration at different energies,  from $\sqrt {s_{NN}} =2.76, 5.02, 7.0$  and 13 TeV \cite{ALICE:2021dhb,ALICE:2017olh,ALICE:2016yta,ALICE:2023sgl}. Particularly interesting is to verify if the introduced model for charm production in pp collisions is also consistent with data at fixed $\sqrt {s_{NN}}$ and for events with different charged-particles rapidity densities, $dN_{ch}/d\eta$.   
\begin{figure}
\includegraphics[scale=0.48]
{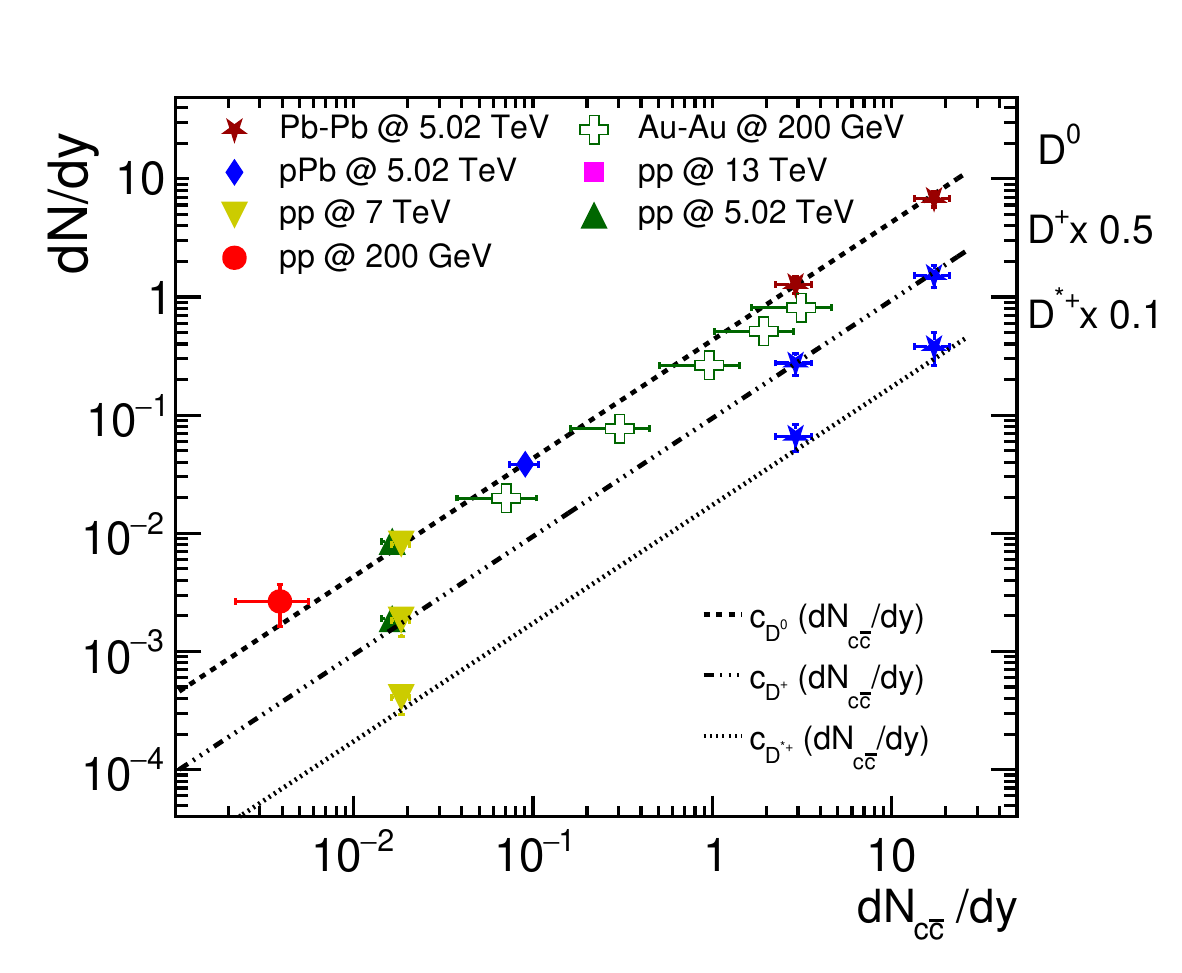} 
\vskip -0.4 cm
\caption{Scaling of 
{$D^0, D^+, D^{*+}$} 
rapidity density with the rapidity density of $c\bar c$ pairs produced in the initial hard scattering. $dN_{c\bar c}/d\eta$ for different systems are  calculated from Eq. \ref{eq:final}. The LHC data are from the ALICE Collaboration \cite{ALICE:2021rxa,ALICE:2019nxm,ALICE:2021mgk,ALICE:2016yta}. The pp data at $\sqrt {s_{NN}}=0.2$  TeV are from the STAR Collaboration \cite{STAR:2014wif}. The Au-Au results are from  Table 1. 
The lines are the SHMc scaling predictions from Eq. \ref{eq:mainAA} at $T_c=156.5$ MeV, see text.}
\label{fig:NCscaling}
\end{figure}

In Fig.~\ref{fig:ccpp} we have summarized the  LHC and RHIC minimum-bias pp data for $N^{pp}_{c\bar c}=\sigma_{c\bar c}^{pp}/\sigma_{inel}^{pp}$ as a function of $dN_{ch}/d\eta$. Also shown is the value for this ratio in pPb collisions {normalized by the $\sqrt \alpha_A$ factor to remove the nuclear modification from the data and make it comparable to pp value}, see discussion above. 

In a rather narrow window, {$3<dN_{ch}/d\eta<18$ }, the $N_{c\bar c}$ is approximately fitted in Fig.~\ref{fig:ccpp} with a linear function of $dN_{ch}/d\eta$.
From Fig. \ref{fig:ccpp} and within the extrapolation one can then extract experimentally unknown  $N^{pp}_{c\bar c}$ values for  $dN_{ch}/d\eta=3.1, 10.5, 22.6$ and 37.8 where the rapidity densities of open charm were measured by the ALICE Collaboration in pp collisions at $\sqrt {s_{NN}}$ = 13 TeV.

With the $N_{c\bar c}$ inputs from   Fig. \ref{fig:ccpp} for pp collisions and using  Eqs. \ref{eq:final} and \ref{eq:mainAA} one compares the model predictions for $D^0, D^+$ and $\Lambda^+_c$ yields with data in Fig. \ref{fig:gc-scaling}.

{The SHMc results for open charm meson yields in minimum-bias pp collisions at   $\sqrt {s_{NN}} = 5.02, 7.0$ and $13$ TeV and predictions for different multiplicity events discussed above are consistent, within experimental uncertainties,  with data. }

{{ 
For the  $\Lambda_c^+$ rapidity density, however, to get  quantitative agreement of model predictions with data one needs to include the contribution of missing charmed-baryon resonances, as discussed above. This is illustrated in Fig.~\ref{fig:gc-scaling} where we compare the SHMc predictions with the  PDG   input (long-dashed lines) and the results obtained after rescaling the contribution of charm baryon densities by 
a phenomenological factor $2.2\pm 0.15$ to account for missing resonances. Following Eq.~\ref{eq:main}, we note that an increase in charmed-baryon density is also modifying by a few percent the rapidity density of open charm mesons.
}}

Furthermore, in high-energy collisions at LHC, the rapidity densities of open charm follow, within uncertainties, 
the power-law scaling with  $dN_{ch}/d\eta$, as already introduced in Fig. \ref{fig:ratioLtoD03}.
{Performing an independent fit to each particle species in  Fig. \ref{fig:gc-scaling} as,
$dN_i/dy=a_i(dN_{ch}/d\eta)^{b_i}$,  one finds that slope parameters within uncertainties are common for all these particles. We get its average value, $\langle b\rangle =0.19\pm 0.022$, with $a_{D^0}=(1.11\pm 0.15)\times 10^{-3}$. The   $a_{D^+}$ and  $a_{\Lambda^+_c}$ proportionality factors are obtained from $a_{D^0}$ by multiplying it by the corresponding SHMc density ratios shown in Figs. \ref{fig:ratioDtoD0}  and \ref{fig:ratioLtoD0}. }
\begin{figure}
\includegraphics[scale=0.45]
{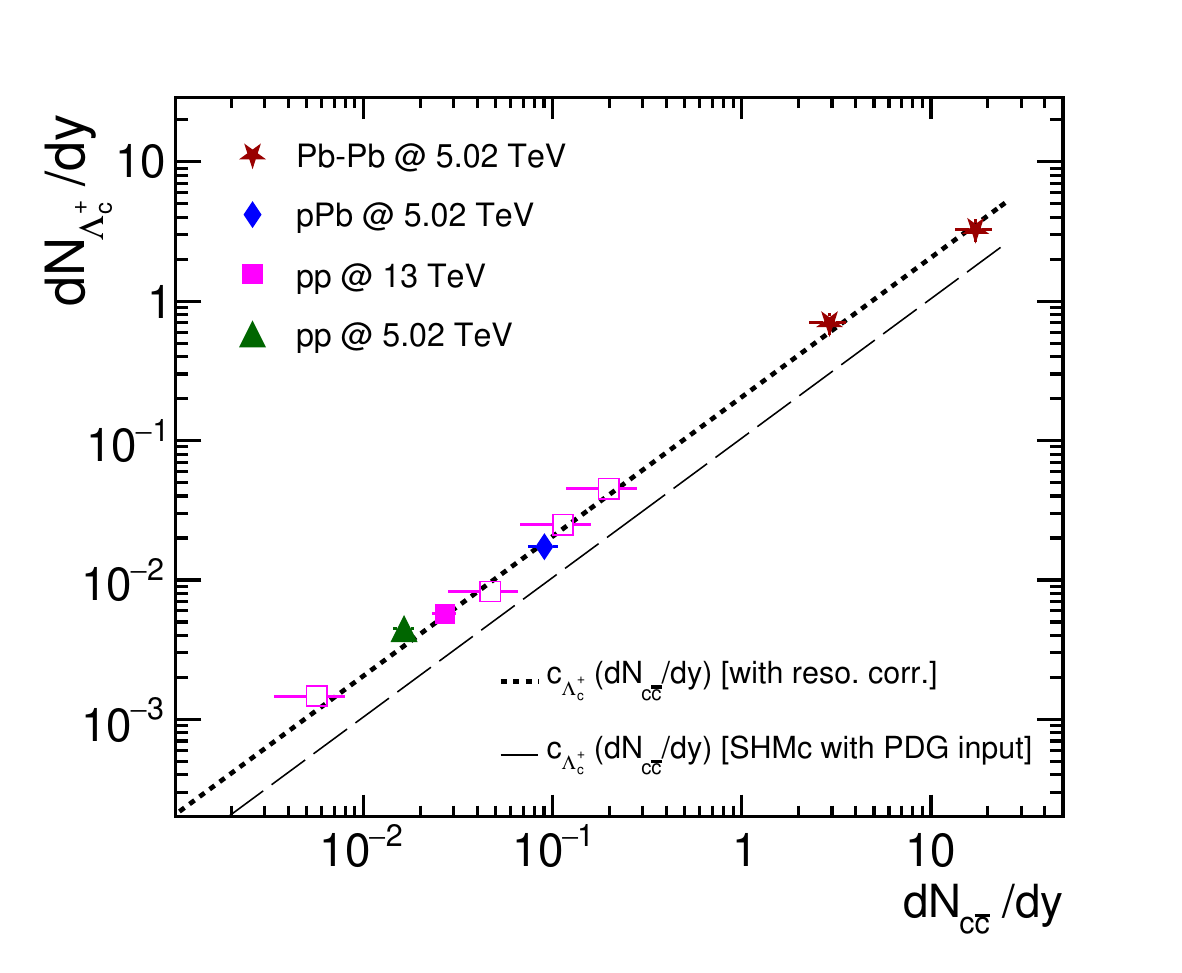}
\vskip -0.4 cm
\caption{Scaling of $\Lambda^{+}_c$ rapidity density with the rapidity density  of $c\bar c$ pairs. $dN_{c\bar c}/d\eta$  for different systems are  calculated from Eq. \ref{eq:final}. The LHC data are from the ALICE Collaboration \cite{ALICE:2021bib,ALICE:2022exq,ALICE:2017thy,ALICE:2023sgl,ALICE:2021npz}. 
The line is the SHMc scaling predictions from Eq. \ref{eq:mainAA} calculated at $T_c=156.5$ MeV with included missing resonances, see text.
{Also shown is the SHMc result calculated with the PDG input for charmed-baryon resonances (long-dashed line). }}
\label{fig:NCscalingL}
\end{figure}

The prediction of the above SHMc in Eq. \ref{eq:mainAA}  is that, in leading order,  the rapidity density of open charm hadrons in high energy pp, pA and AA collisions should closely follow the proportional scaling with the rapidity density of the number of  $c\bar c$ pairs.

In Figs.  \ref{fig:NCscaling} and \ref{fig:NCscalingL}
we summarize data on $D^0, D^+, D^{*+}$ mesons and $\Lambda_c^+$ baryon rapidity density at different collision energies and colliding systems.  With  the method  introduced in Eq. \ref{eq:final} 
to calculate $N_{c\bar c}$,  all    data  indeed follow within uncertainties  the  proportional scaling with $N_{c\bar c}$,  as  $dN_i/dy=c_i N_{c\bar c}$.
The proportionality coefficients,   $c_{D^0}=0.43$ and $c_{\Lambda_c^+}=0.205$ in Figs. \ref{fig:NCscaling} and \ref{fig:NCscalingL},   are  calculated in the SHMc 
from the corresponding  density ratio in 
Eq. \ref{eq:mainAA}  at $T_c=156.5$ MeV,  including the contribution of missing charmed baryon resonances, as discussed above. 
{ The SHMc results for $dN_{\Lambda_c^+}/dy$   obtained  with the PDG resonance input shown in Fig. \ref{fig:NCscalingL} underpredict the scaling coefficient $c_{\Lambda_c^+}$ in Eq. 8 by a factor of $1.97\pm 0.14$. }   

Considering the yet incomplete knowledge of the charm mass spectrum, we have also fitted
 slopes to 
 results in   Figs. \ref{fig:NCscaling} and \ref{fig:NCscalingL}
   as:  $c_{D^0}= 0.45 \pm  0.046$  and $c_{\Lambda^+_c}=0.203 \pm  0.053$,  which contain the above  SHMc predicted values 
   {if in the case of $\Lambda_c^+$ one includes the contribution of missing resonances, as discussed above.  }

From the observed scaling in Figs. \ref{fig:NCscaling} and \ref{fig:NCscalingL},  one concludes that the SHMc provides a good description of charm quark fragmentation to open charm hadron species carrying charm quantum-number  $|c|=1$, independent of collision energy and colliding systems. Furthermore, from Figs. \ref{fig:gc-scaling}-\ref{fig:NCscalingL}, one concludes that   the rapidity density of charm-anticharm quark pairs produced in the initial state, $dN_{c\bar c}/dy$, exhibits a power-law scaling with the observed 
charged-particle pseudo-rapidity density,  $dN_{ch}/d\eta$.

\section{Summary and Conclusions}
\label{summary}

Considering recent data on open charm hadron production yields in proton-proton, proton-nucleus and heavy ion collisions at the LHC and top RHIC energy, we have analyzed their properties and interpretation within the Statistical Hadronization Model for charm (SHMc).  We have extended and used this model to quantify data on open charm production in minimum-bias and high-multiplicity pp collisions 
 at different energies and discuss their link to heavy ion collisions.

We have focused on the rapidity density, $dN_i/dy$,  data of $D^0, D^+, D^{*+}, D_s^+$ mesons and $\Lambda_c^+$ baryons in  Pb-Pb at $\sqrt { s_{NN}}=5.02$ TeV by the ALICE  experiment at LHC,  and on $D^0$ meson production in Au-Au at $\sqrt {s_{NN}}= 0.2$ TeV by the STAR experiment at RHIC. Furthermore,  we have linked heavy ion with proton-proton (pp) minimum-bias data measured at $\sqrt {s_{NN}} = 5.02, 7.0$ and 13 TeV by ALICE and at $\sqrt { s_{NN}}=0.2$ TeV by STAR experiments.   At $\sqrt {s_{NN}}= 13$ TeV the ALICE  pp data also concerned events with different charged-particle pseudo-rapidity densities, $dN_{ch}/d\eta$.

The present analysis demonstrates that most currently available data on different ratios of open charm rapidity densities in high-energy collisions are independent of collision energy and system size, as expected in the SHMc. On the quantitative level, the yield ratios of $D^+, D^{*+}$ and $D^0$ are consistent with SHMc with PDG  input. An exception are ratios involving charmed baryons, in particular, the  $\Lambda^+_c/D^0$ ratio is larger by a factor of $2.2$, which is attributed to missing resonances in the charm baryon mass spectrum. The observed suppression of $D^+_s/D^0$ ratio by a factor of {nearly} two relative to SHMc predictions for pp and pPb collisions requires further studies.  We note, however, that data on $D^+_s/D^0$  from Au-Au and Pb-Pb collisions agree with SHMc predictions without any suppression, see Fig. \ref{fig:ratioDtoD0}. Clearly,  it would be very valuable to have a precision measurement of the $D^+_s/D^0$ ratio as a function of $dN_{ch}/d\eta$ from pp to Pb-Pb collisions.

We have further demonstrated that, according to SHMc predictions,  the rapidity densities $dN_i/dy$ of open charm hadrons with charm quantum-number $|c|=1$, should scale proportional to the number of initially produced charm quark pairs,   $dN_{c\bar c}/dy$ in high energy pp and AA collisions.  Indeed,  $dN_{D^0}/dy$  data in pp, pPb and Pb-Pb collisions at the LHC  and top RHIC energies 
are following the predicted scaling with the slope quantified by the SHMc at the QCD chiral crossover temperature.
Such a scaling was also quantified for $D^{*+}, D^+$ and $\Lambda^+_c$ rapidity density. However, for the case of $D^+_s$ the $N_{c\bar c}$ scaling is violated as pp 
collisions exhibit a different scaling from that observed in Au-Au and Pb-Pb collisions, see the discussion above.

An interesting further production systematics of the rapidity density of open charm hadrons found in this study is their scaling with  $dN_{ch}/d\eta$. The pp, pPb and Pb-Pb rapidity density data at LHC from $\sqrt {s_{NN}}=5.02, 7.0$ and 13 TeV for open charm species $i=D^0, D^+, D^{*+}, \Lambda^+_c$  follow, within uncertainties, a power-law scaling, $dN_i/dy=a_i(dN_{ch}/d\eta)^b$, where the power $b\simeq 0.2$  is common for all single charmed hadrons. Such scaling is also found in  $dN_{D^0}/dy$  data at RHIC from minimum-bias pp to central Au-Au collisions at $\sqrt {s_{NN}}=0.2$ TeV.

{ The convergence of model predictions and data for  $D^0,D^+$ and $D^{*+}$ open charm   meson  yields,  shown  in Figs. 1, 6 and 7,
is further evidence that the main concept of SHMc, proposed in 
~\cite{Braun-Munzinger:2000eyl,BraunMunzinger:2000px,Andronic:2021erx} 
is realized. This concept assumes the thermalization of charm quarks produced in the initial state and their subsequent hadronization at the QCD interface at $T_c\simeq 157$ MeV, under the constraint of 
conservation of the number of $c\bar c$ pairs from the initial partonic to the final hadronic state.
Furthermore, the above observed agreement of model predictions and open charm meson data also implies that  SHMc provides a good description of charm quark fragmentation into open charm mesons. 
The production systematics of $\Lambda_c^+$ baryon qualitatively follow the SHMc predictions. However, the slope coefficients for $dN_{\Lambda_c^+}/dy$  shown in Figs. 6 and 8 differ in the SHMc with PDG input  at $T_c$ by a factor of $\sim 2$  from the data. Following 
the recent LQCD results we have attributed the above differences to missing resonances in the charmed baryon sector of the  PDG  mass spectrum. Rescaling the contributions of charmed-baryon density by a factor of $2.2\pm 0.15$ in the  SHMc with the PDG input results in the shift of $dN_{\Lambda_c^+}/dy$ value to that expected in the data.  }


Considering the somewhat ad-doc correction for $\Lambda^+_c$ baryons made above it would be extremely important to shed more light on this situation by experimentally searching for missing charmed baryon resonances. Note that higher-lying charmed baryons increase via strong decays to $\Lambda^+_c$ yield by about a factor of two. Consequently, our understanding of charm production will benefit significantly from an improved understanding of this decay cascade. In particular, it will allow for precise determination of the freezeout temperature of charmed hadrons to distinguish if they are produced at the chiral crossover, as assumed in these studies,  or rather at higher temperatures as suggested e.g. in  
\cite{He:2019tik}.

The concept of the thermal origin of heavy flavour production in high energy pp, pA and AA collisions and their scaling,  introduced above,   is not restricted 
  to charm-quark bound-states. It can also be extended to bottom-hadron production with similar systematics and physics concepts, provided bottom quarks are also thermalizing in high energy collisions \cite{Andronic:2022ucg}.   

\section*{Acknowledgments} 
This work was performed within and is supported by the DFG Collaborative Research Centre "SFB 1225 (ISOQUANT)". K.R. acknowledges the Polish National Science Centre (NCN) support under OPUS Grant No. 2022/45/B/ST2/01527 and of the Polish Ministry of Science and Higher Education. K.R. also acknowledges the valuable comments of Frithjof Karsch, Chihiro Sasaki and Sipaz Sharma. N.S.  acknowledges the discussions with Boris Hippolyte and Lokesh Kumar. We are also thankful for the stimulating discussions and comments from Anton Andronic, L. McLerran and Nu Xu.
\bibliography{sample}

\end{document}